\documentclass[conference]{IEEEtran}
\IEEEoverridecommandlockouts
\usepackage[noadjust]{cite}
\usepackage{amsmath}
\usepackage{array}
\usepackage{mdwmath}
\usepackage{amssymb}
\usepackage{mathtools}
\usepackage{algorithm}
\usepackage[noend]{algpseudocode}
\usepackage{eqparbox}
\usepackage{stfloats}
\usepackage{url}
\usepackage[pdftex]{graphicx}
\graphicspath{{../pdf/}{../jpeg/}}
\DeclareGraphicsExtensions{.pdf,.jpeg,.png}
\usepackage{epstopdf}
\usepackage{amsfonts}

\usepackage{color}

\usepackage{tikz}
\usepackage{pgfplots}
\usepackage{subcaption}
\usetikzlibrary{spy}

\usepackage{pgfplots}
\usepackage{caption}
\usepackage{tikz,pgfplots}
\pgfplotsset{compat=1.11}

\newtheorem{lemma}{Lemma}
\newtheorem{theorem}{Theorem}

\newtheorem{corollary}{Corollary}

\begin{document}

\title{Reconfigurable Intelligent Surfaces Assisted Communication Under Different CSI Assumptions}
\author{\IEEEauthorblockN{Bayan Al-Nahhas\IEEEauthorrefmark{1}, Qurrat-Ul-Ain Nadeem\IEEEauthorrefmark{1}, Aryan Kaushik, and Anas Chaaban\IEEEauthorrefmark{1}\IEEEauthorrefmark{2}}
\IEEEauthorblockA{\IEEEauthorrefmark{1} School of Engineering, University of British Columbia, Kelowna, Canada. \\ Email: \{bayan.alnahhas, qurrat.nadeem, anas.chaaban\}@ubc.ca}
\IEEEauthorblockA{\IEEEauthorrefmark{2} School of Engineering and Informatics, University of Sussex, Brighton, UK.  Email: aryan.kaushik@sussex.ac.uk}
}

\maketitle
\begin{abstract}
This work studies the net sum-rate performance of a distributed reconfigurable intelligent surfaces (RISs)-assisted multi-user multiple-input-single-output (MISO) downlink communication system under imperfect instantaneous-channel state information (I-CSI) to implement precoding at the base station (BS) and statistical-CSI (S-CSI) to design the RISs phase-shifts. Two channel estimation (CE) protocols are considered for I-CSI acquisition: (i) a full CE protocol that estimates all direct and RISs-assisted channels over multiple training sub-phases, and (ii) a low-overhead direct estimation (DE) protocol that estimates the end-to-end channel in a single sub-phase. We derive the asymptotic equivalents of signal-to-interference-plus-noise ratio (SINR) and  ergodic net sum-rate under  both protocols for given RISs phase-shifts, which are then optimized based on S-CSI. Simulation results reveal that the low-complexity DE protocol yields better net sum-rate  than the full CE protocol when used to obtain CSI for precoding. A benchmark full I-CSI based RISs design is also outlined and shown to yield higher SINR  but lower net sum-rate than the S-CSI based RISs design. 

\end{abstract}


\section{Introduction}

Deploying reconfigurable intelligent surfaces (RISs) on different structures in the environment has emerged as a transformative solution to   customize the propagation of radio waves through controlled reflections  \cite{Wu1}. Specifically, an RIS constitutes of a large number of  low-cost passive reflecting elements that  induce  phase shifts onto the incident signals, that can be  smartly tuned to realize desired performance objectives  \cite{Wu1, annie}. While significant performance gains have been shown using a single RIS, the use of  distributed RISs  can  more effectively enhance coverage,  enable communication when multiple direct links are weak, improve the channel rank, and increase the spectral and energy efficiency gains, and have been the focus of some recent works \cite{sumrate_dis, maxmin_dis2, maxmin_dis3}.

Most of the current literature on RIS-assisted  systems assumes perfect channel state information (CSI) of all links to be available for beamforming design, which is  impractical  given channel estimation (CE) is challenging in RIS-assisted systems. In this context,  least-square and minimum mean squared error (MMSE) channel estimates of the direct base station (BS)-user links and RIS-assisted links have been derived in \cite{annie_OJ}, using a  pilot training based CE protocol requiring $N+1$ training sub-phases, where $N$ is the number of RIS elements.   Some works  develop lower overhead  protocols by grouping RIS elements,  exploiting  the static  nature of the BS-RIS channel \cite{hiba_CE}, or utilizing the correlation between the channels of different users  \cite{CER1}, but impose a training overhead that  still grows  large with $N$. Moreover, optimizing the RISs based on instantaneous CSI (I-CSI) at the pace of a fast-fading channel significantly increases the system complexity.



 To mitigate these challenges, the authors in \cite{IRSstat, maxmin_dis3} design the RIS parameters using  statistical CSI (S-CSI) without requiring the I-CSI of individual BS-RIS and RIS-users channels. The only I-CSI then needed is of the aggregate end-to-end channel to design precoding at the BSs. Such designs not only eliminate the training overhead requirements associated with the  estimation of individual  links, but  also  relax  the need for frequently reconfiguring the RISs. The works that study such designs (e.g.  \cite{IRSstat, maxmin_dis3}) assume the I-CSI of the aggregate channel  to  be perfectly known at the BS, which is impractical. Moreover, they do not compare the performance of S-CSI and  I-CSI based RIS designs in terms of training overhead, and net sum-rate. These constitute the main questions of this paper.
 
 Specifically,  we study a distributed RISs-assisted multi-user multiple-input single-output (MISO) communication system,  considering the scenarios of (i) full imperfect I-CSI \cite{hiba_CE} versus aggregate imperfect I-CSI \cite{IRSstat} at the BS to implement precoding, and (ii) I-CSI versus S-CSI availability to design the RISs phase-shifts. To this end, we utilize the  MMSE-discrete Fourier transform (DFT) CE protocol from \cite{hiba_CE} to obtain  full I-CSI.  Recognizing its large overhead in terms of the required CE sub-phases, we consider a second direct estimation (DE)  protocol, which estimates each  end-to-end BS-user channel in a single sub-phase for given RISs  phase shifts. We  analytically study the ergodic net sum-rate  achievable under maximum ratio transmission (MRT) precoding implemented  at the BS using imperfect I-CSI of the end-to-end channel under both CE protocols. The derived expressions are utilized to optimize the RISs phases based on S-CSI. Later we study the net-sum rate performance when RISs phase shifts are designed based on full imperfect I-CSI. The results show that DE yields higher net sum-rate than the MMSE-DFT CE protocol,  when combined with the S-CSI based RISs design, especially for large system sizes. Moreover the full I-CSI based RISs design yields high signal-to-interference-plus-noise ratio (SINR)  but its net sum-rate performance is worse than that of the DE+S-CSI based scheme.
\vspace{-0.1 in}
\section{System Model and Problem Formulation}
\begin{figure}[t!]
\centering
\includegraphics[scale=.28]{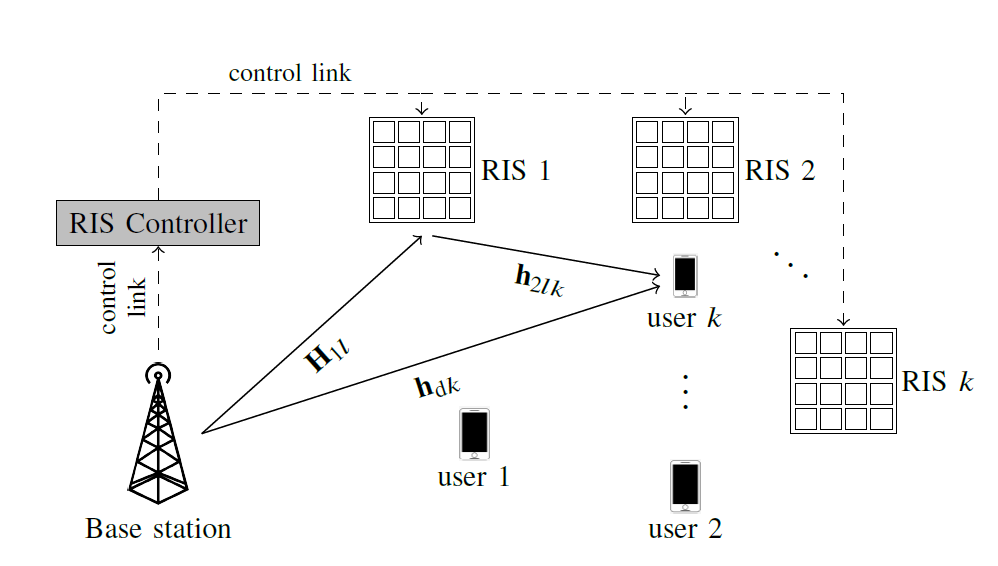}
\caption{Distributed RISs-assisted communication model. }
\label{LIS_model}
\end{figure}

As shown in Fig. \ref{LIS_model}, we consider a BS equipped with $M$ antennas communicating with $K$ single antenna users, with the assistance of  $L$ RISs composed of $N$  reflecting elements each that are connected to the BS via a control link.
\subsection{Signal and Channel Models}
\label{sysmodel}
 The BS wants to send information at rate $R_k$ to user $k$. To this end, the BS constructs codewords with symbols $s_k\sim \mathcal{CN}(0,1)$, and combines them into the transmit (Tx) signal vector $\mathbf{x}$ given as $\mathbf{x}=\sum_{k=1}^{K} \sqrt{p_{k}} \mathbf{g}_{k} s_{k}$, with $\mathbf{g}_{k} \in \mathbb{C}^{M\times 1}$ and $p_{k}>0$ being the precoding vector and signal power for user $k$ respectively. The Tx signal  satisfies the power constraint $ \mathbb{E}[||\mathbf{x}||^{2}]=\mathbb{E}[\text{tr}(\mathbf{P} \mathbf{G}^{H} \mathbf{G})] \leq  P_{max}$, where   $P_{max}$ is the  Tx power budget, $\mathbf{P}=\text{diag}(p_{1}, \dots, p_{K})$ and $\mathbf{G}=[\mathbf{g}_{1}, \dots,\mathbf{g}_{K}]$. The received  signal $y_{k}$ at user $k$ is given as  \vspace{-.05in}
\begin{align}
\label{y_k}
&y_{k}= \mathbf{h}_k^{H} \mathbf{x}+n_k,
\end{align}
where $n_{k}\sim \mathcal{CN}(0,\sigma^2)$ is the noise, and  $\mathbf{h}_k$ is the channel between the BS and user $k$ given as \vspace{-.05in}
\begin{align}
\label{eq_ch}
&\mathbf{h}_{k}= \mathbf{h}_{dk}+\sum_{l=1}^L\mathbf{H}_{1l}\boldsymbol{\Theta}_l \mathbf{h}_{2lk},
\end{align} 
where $\mathbf{H}_{1l}\in \mathbb{C}^{M\times N}$ is the channel between  RIS $l$ and the BS, and $\mathbf{h}_{2lk} \in \mathbb{C}^{N\times 1}$ and  $\mathbf{h}_{dk}\in \mathbb{C}^{M\times 1}$  are the  channels between user $k$ and  RIS $l$, and  user $k$ and the BS respectively. Also $\boldsymbol{\Theta}_l=\text{diag}(\phi_{l1},\ldots,\phi_{lN})$ represents the response of  RIS $l$, where $\phi_{ln}=\alpha_{ln} e^{j\theta_{ln}}$, and $\theta_{ln}\in[0,2\pi] $ and $\alpha_{ln}=1$ are the phase and amplitude of reflection coefficient of  element $n$.

Each BS-RIS channel is considered to be line-of-sight (LoS) dominated \cite{ LoS1,  annie, hiba_CE}, since the LoS path between the BS and RIS can be guaranteed through  appropriate deployment of the RISs, and the non-LoS paths are much weaker than the LoS path at mmWave frequencies.   We consider a uniform rectangular array of $N=N_1\times N_2$  elements  at each RIS with inter-element spacing of $d_{RIS}^{(1)}$ and $d_{RIS}^{(2)}$ along the two directions, and a uniform linear array  with inter-antenna spacing of $d_{BS}$ at the BS. Under the spherical wave model, we obtain $[\mathbf{H}_{1l}]_{m,n}=\sqrt{\beta_{1l}}\exp\left(j \frac{2\pi}{\lambda} \bar{l}_{(m),(n_1,n_2)_l}\right)$,  where $n=(n_1-1)N_2+n_2$, $n_1=1,\dots, N_1$, $n_2=1, \dots, N_2$,  $\beta_{1l}$ is the path loss factor, and $\bar{l}_{(m),(n_1,n_2)_l}$ is the path length between  BS antenna $m$ and RIS $l$'s element $(n_1,n_2)$ \cite{losura}.


For  $\mathbf{h}_{2lk}$ and  $\mathbf{h}_{dk}$,  we consider Rician  fading models as
\begin{align}
\label{ch111}
&\mathbf{h}_{2lk}= \mathbf{h}^{\rm n} _{2lk}+\bar{\mathbf{h}}_{2lk}, \hspace{.12in} \mathbf{h}_{dk}= \mathbf{h}^{\rm n}_{dk}+\bar{\mathbf{h}}_{dk}
\end{align}
where $\mathbf{h}_{2lk}^{\rm n}=\sqrt{\frac{1}{\kappa_{2lk}+1}}\mathbf{z}_{2lk}$ and $\mathbf{h}_{dk}^{\rm n}=\sqrt{\frac{1}{\kappa_{dk}+1}}\mathbf{z}_{dk}$ are the NLoS components,  where $\mathbf{z}_{2lk} \sim \mathcal{CN}(0,\beta_{2lk} \mathbf{I}_N)$ and $\mathbf{z}_{dk}\sim\mathcal{CN}(0,\beta_{dk} \mathbf{I}_M)$. The LoS channel components  $\bar{\mathbf{h}}_{2lk}$ and $\bar{\mathbf{h}}_{dk}$ are modeled as $\bar{\mathbf{h}}_{dk}=\sqrt{\frac{{\beta}_{dk}\kappa_{dk}}{\kappa_{dk}+1}}[1, e^{j2\pi d_{BS} \cos (\phi_{dk})},  \dots  , e^{j2\pi d_{BS} (M-1) \cos (\phi_{dk})}]$ and  $\bar{\mathbf{h}}_{2lk}=\sqrt{\frac{ {\beta}_{2lk} \kappa_{2lk}}{\kappa_{2lk}+1}} [\mathbf{b}_{2lkz} \otimes \mathbf{b}_{2lkx}]$, where $\mathbf{b}_{2lkz}=[1, e^{j2\pi d^{(1)}_{RIS} \cos (\phi_{2lk})},  \dots  , e^{j2\pi d_{RIS}^{(1)} (N_1-1) \cos (\phi_{2lk})}]$,  $\mathbf{b}_{2lkx}=[1, e^{j2\pi d^{(2)}_{RIS} \cos (\phi_{2lk})},  \dots  , e^{j2\pi d_{RIS}^{(2)} (N_2-1) \cos (\phi_{2lk})}]$, and $\phi_{dk}$ and $\phi_{2lk}$ are the angles of departure (AoD) of the wave-vector from the BS and RIS $l$ to user $k$ respectively. Moreover, $\beta_{dk}$ and $\beta_{2lk}$ are the path loss factors, and  $\kappa_{dk}$ and $\kappa_{2lk}$ are the Rician factors for the BS-user $k$ link and RIS $l$-user $k$ link respectively. The overall channel between user $k$ and the BS can be written as 
\begin{align}
\label{eq_ch1_RIC}
\mathbf{h}_{k}=\mathbf{h}^{\rm n}_{dk}+\bar{\mathbf{h}}_{dk}+\sum_{l=1}^L \mathbf{H}_{1l}\boldsymbol{\Theta}_l \mathbf{h}^{\rm n}_{2lk}+\sum_{l=1}^L \mathbf{H}_{1l}\boldsymbol{\Theta}_l \bar{\mathbf{h}}_{2lk}
\end{align} 
which is statistically equivalent to a correlated Rician channel as $\mathbf{h}_{k}=\bar{\mathbf{h}}_{dk}+\sum_{l=1}^L \mathbf{H}_{1l}\boldsymbol{\Theta}_l \bar{\mathbf{h}}_{2lk}+\mathbf{A}^{{1/2}}_{k}\mathbf{z}_{k}$, where $\mathbf{z}_{k}\sim \mathcal{CN}(\mathbf{0}, \mathbf{I}_{M})$ and $\mathbf{A}_{k}=\frac{ \beta_{dk}}{\kappa_{dk}+1}\mathbf{I}_{M}+\sum_{l=1}^L \frac{\beta_{2lk}}{\kappa_{2lk}+1}\mathbf{H}_{1l} \mathbf{H}_{1l}^H$.

\subsection{CE Protocols}

We consider two protocols  to obtain  I-CSI for precoding.


\subsubsection{MMSE-DFT CE Protocol}  The MMSE-DFT CE protocol from \cite{hiba_CE}  constructs the estimate of the aggregate channel $\mathbf{h}_k$ by estimating $\mathbf{h}_{2lk}$'s and $\mathbf{h}_{dk}$ over $S$  CE sub-phases. In this protocol, RIS $l$ applies the reflect beamforming matrix $\boldsymbol{\Theta}_{ls} = \text{diag}(\phi_{ls1}, \dots, \phi_{lsN}) \in \mathbb{C}^{N \times N}$ in sub-phase $s \in \{1,\dots, S  \}$, resulting in the RISs training matrix  
\begin{align}
\mathbf{V}_{tr}=\begin{bmatrix} 1 & \mathbf{v}_{11}^T & \dots & \mathbf{v}_{L1}^T\\
	\vdots& 	\vdots & &\vdots\\ 
      1 & \mathbf{v}_{1S}^T & \dots & \mathbf{v}_{LS}^T
  \end{bmatrix} \in \mathbb{C}^{S\times (NL+1)} 
  \end{align}
   where $\mathbf{v}_{ls}=\text{diag}(\boldsymbol{\Theta}_{ls}) $.     The optimal $\mathbf{V}_{tr}$ that minimizes the CE error  is derived to be the $NL+1$ leading columns of a DFT matrix as $[\mathbf{V}_{tr}]_{s,n}= e^{-j2\pi(n-1)(s-1)/S }$  in \cite{hiba_CE}.  This protocol  exploits the LoS nature of the BS-RISs channels to reduce the number of training sub-phases to $S=\frac{NL}{M}+1$  as compared to other CE protocols in \cite{ annie_OJ,CER1}. Here we extend its results to out setup, and derive the estimates as follows.


\begin{lemma}\label{L11_RIC}
The MMSE estimate of $\mathbf{h}_{k}$ in \eqref{eq_ch1_RIC} under the MMSE-DFT CE protocol is given as  \vspace{-.12in}
\begin{align}
\label{est_corr11ric}
\hat{\mathbf{h}}_{k}= \hat{\mathbf{h}}^{\rm n}_{dk}+\bar{\mathbf{h}}_{dk}+\sum_{l=1}^L \mathbf{H}_{1l}\boldsymbol{\Theta}_l \hat{\mathbf{h}}^{\rm n}_{2lk}+\sum_{l=1}^L \mathbf{H}_{1l}\boldsymbol{\Theta}_l \bar{\mathbf{h}}_{2lk},
\end{align}
where $\hat{\mathbf{h}}_{dk}^n$ and $\hat{\mathbf{h}}_{2lk}^n$ are the MMSE estimates of $\mathbf{h}^{\rm n}_{dk}$ and $\mathbf{h}^{\rm n}_{2lk}$  given as $\hat{\mathbf{h}}^{\rm n}_{dk}=\frac{\beta^{\rm n}_{dk}}{\beta^{\rm n}_{dk}+\frac{1}{S \rho_{p} \tau_S}}(\tilde{\mathbf{r}}^{tr}_{0k}-\bar{\mathbf{h}}_{dk})$ and $ \hat{\mathbf{h}}^{\rm n}_{2lk}=\frac{\beta^{\rm n}_{2lk}}{\beta^{\rm n}_{2lk}+\frac{1}{S \rho_{p} \tau_S M \beta_{1l}} }(\tilde{\mathbf{r}}^{tr}_{lk}-\bar{\mathbf{h}}_{2lk})$
where $\beta^{\rm n}_{dk}=\frac{\beta_{dk}}{\kappa_{dk}+1}$,  $\beta^{\rm n}_{2lk}=\frac{\beta_{2lk}}{\kappa_{2lk}+1}$, $\rho_p$ is the training SNR, $\tau_S$ is the length of each training sub-phase, and the observation vectors $\tilde{\mathbf{r}}_{0k}^{tr}$ and $\tilde{\mathbf{r}}_{lk}^{tr}$  are given as $\tilde{\mathbf{r}}^{tr}_{0k}=\mathbf{h}^{\rm n}_{dk}+\bar{\mathbf{h}}_{dk}+\frac{1}{S} (\mathbf{v}_1^{tr} \otimes \mathbf{I}_{M})^H \mathbf{{n}}^{tr}_{k}$ and $ \tilde{\mathbf{r}}^{tr}_{lk}=\mathbf{h}^{\rm n}_{2lk}+\bar{\mathbf{h}}_{2lk}+\frac{1}{SM\beta_{1l}}\bar{\mathbf{H}}_{1l}^H(\mathbf{V}_l^{tr} \otimes \mathbf{I}_{M})^H \mathbf{{n}}^{tr}_{k}$, where  $\mathbf{n}_k^{tr} \in \mathbb{C}^{MS \times 1}$ is the received noise across $S$ CE sub-phases,  $\mathbf{v}_1^{tr}$ is the first $S \times 1$ column of  $\mathbf{V}_{tr}$, and $\mathbf{V}_l^{tr}\in \mathbb{C}^{S \times N}$ comprises of the $N(l-1)+2$ to $Nl+1$ columns of $\mathbf{V}_{tr}$  for $ l=1,\dots, L$. Moreover $\bar{\mathbf{H}}_{1l}= \text{diag}(\bar{\mathbf{h}}_{1l1}, \dots,  \bar{\mathbf{h}}_{1lN}) \hspace{-.04in} \in \hspace{-.04in} \mathbb{C}^{MN \times N}$\hspace{-.04in},  where $\bar{\mathbf{h}}_{1ln}$ is the $n$th column of  $\mathbf{H}_{1l}$, known due to its LoS nature.
\end{lemma}
\begin{IEEEproof}
The proof follows by applying the definition of MMSE estimate on the observation vectors \cite[Sec. III]{hiba_CE}. 
\end{IEEEproof}

Note that the CE error $\tilde{\mathbf{h}}^{\rm n}_{dk}=\mathbf{h}^{\rm n}_{dk}-\hat{\mathbf{h}}^{\rm n}_{dk}$, which is also Gaussian, is independent of $\hat{\mathbf{h}}^{\rm n}_{dk}$. A similar discussion applies to  $\tilde{\mathbf{h}}^{\rm n}_{2lk}=\mathbf{h}^{\rm n}_{2lk}-\hat{\mathbf{h}}^{\rm n}_{2lk}$. Using these results, $\hat{\mathbf{h}}_{k}$ is statistically equivalent to a correlated Rician channel as follows.

\begin{lemma} \label{L1_RIC} The  estimate $\hat{\mathbf{h}}_{k}$ in \eqref{est_corr11ric} can be represented as \vspace{-.05in}
\begin{align}
\label{est_corr_RIC}
& \hat{\mathbf{h}}_{k}= \bar{\mathbf{h}}_{dk}+\sum_{l=1}^L \mathbf{H}_{1l}\boldsymbol{\Theta}_l \bar{\mathbf{h}}_{2lk}+\mathbf{C}^{{1/2}}_{k}\mathbf{q}_{k},
\end{align}
where $\mathbf{q}_{k}\sim \mathcal{CN}(0,\mathbf{I}_{M})$ and $\mathbf{C}_{k}=\frac{\beta_{dk}^{{n}^{2}}}{\beta^{\rm n}_{dk}+\frac{1}{S\rho_{p} \tau_S}} \mathbf{I}_M+\sum_{l=1}^L \frac{\beta_{2lk}^{{n}^{2}}}{\beta^{\rm n}_{2lk}+\frac{1}{S\rho_{p} \tau_S M \beta_{1l}}} \mathbf{H}_{1l}  \mathbf{H}_{1l}^H$.
\end{lemma}

This  protocol accurately estimates all channels  at the expense of a large training overhead which may compromise the  net sum-rate. Next, we present a lower overhead CE protocol.

\subsubsection{DE Protocol}


In the DE scheme, instead of estimating the individual channels $\mathbf{h}_{dk}$'s and $\mathbf{h}_{2lk}$'s, the BS directly estimates   the aggregate channel $\mathbf{h}_{k}=\mathbf{h}_{dk}+\sum_{l=1}^L \mathbf{H}_{1l}\boldsymbol{\Theta}_l\mathbf{h}_{2lk}$ for given  $\boldsymbol{\Theta}_l$'s in a single sub-phase.  The  estimate of $\mathbf{h}_k$ will therefore depend on the choice of RIS phase shifts. We propose that $\boldsymbol{\Theta}_l$'s are computed based on S-CSI  (as discussed in the next section) at the start of a time-frame over which the channel statistics stay constant. Then in each coherence block, the BS only   estimates $\mathbf{h}_{k}$, for these given $\boldsymbol{\Theta}_l$'s as follows.

\begin{lemma} \label{L4ric}  The MMSE estimate of $\mathbf{h}_{k}$ under  DE  is  \vspace{-.05in}
\begin{align}
\label{est_deric}
& \hat{\mathbf{h}}_{k}= \bar{\mathbf{h}}_{dk}+\sum_{l=1}^L\mathbf{H}_{1l}\boldsymbol{\Theta}_l\bar{\mathbf{h}}_{2lk}+\mathbf{R}_{k} \mathbf{Q}_{k} \tilde{\mathbf{y}}_{k}^{tr},
\end{align} 
where $\tilde{\mathbf{y}}_{k}^{tr}=\mathbf{y}_{k}^{tr}-\bar{\mathbf{h}}_{dk}-\sum_{l=1}^L\mathbf{H}_{1l}\boldsymbol{\Theta}_l\bar{\mathbf{h}}_{2lk}$, $\mathbf{y}_{k}^{tr}=\mathbf{h}_{dk}+\sum_{l=1}^L\mathbf{H}_{1l}\boldsymbol{\Theta}_l{\mathbf{h}}_{2lk}+\mathbf{n}_k^{tr}$, $\mathbf{n}^{tr}_{k}\sim \mathcal{CN}(\mathbf{0}, \frac{1}{\rho_p \tau_S}\mathbf{I}_M)$, $\mathbf{Q}_{k}=\left(\mathbf{R}_{k}+\frac{ \mathbf{I}_M}{\rho_{p} \tau_S} \right)^{-1}$, and $\mathbf{R}_{k}=\beta^{\rm n}_{dk}\mathbf{I}_M +\sum_{l=1}^L\beta^{\rm n}_{2lk} \mathbf{H}_{1l} \mathbf{H}_{1l}^H$. The channel estimate $\hat{\mathbf{h}}_{k}$  is statistically equivalent to $ \hat{\mathbf{h}}_{k}= \mathbf{C}_{k}^{{1/2}}\mathbf{q}_{k}$, where $\mathbf{q}_{k}\sim \mathcal{CN}(0,\mathbf{I}_{M})$ and $\mathbf{C}_{k}=\mathbf{R}_{k} \mathbf{Q}_{k}\mathbf{R}_{k} $.
\end{lemma}
 \begin{IEEEproof}
The proof follows by estimating  $\mathbf{h}_{k}^{\rm n}=\mathbf{h}_{dk}^{\rm n}+\sum_{l=1}^L\mathbf{H}_{1l}\boldsymbol{\Theta}_l{\mathbf{h}}^{\rm n}_{2lk}$  using $\mathbf{y}_{k}^{tr}-\bar{\mathbf{h}}_{dk}-\sum_{l=1}^L\mathbf{H}_{1l}\boldsymbol{\Theta}_l\bar{\mathbf{h}}_{2lk}$.
\end{IEEEproof}

While this protocol does not provide full I-CSI of the individual RIS-assisted links, it provides enough information ($\hat{\mathbf{h}}_k$) to implement precoding at the BS. It also saves the large training overhead associated with $S\geq \frac{NL}{M}+1$ sub-phases  to obtain the full I-CSI under MMSE-DFT protocol, and requires only a single training sub-phase.  The downside is that  the estimate in \eqref{est_deric} can not be used to design the RISs phases instantaneously. However, if the  phase-shifts are designed using  S-CSI as we do next, then DE is a  desirable scheme because the BS can use \eqref{est_deric}  instead of \eqref{est_corr11ric} for precoding.
\subsection{Precoding and Achievable Rate}
 The  estimate $\hat{\mathbf{h}}_k$ obtained from both CE methods is used to implement MRT precoding at the BS, which is known to be asymptotically optimal for a MISO broadcast channel as $M$  grows large and has a smaller computational complexity compared to zero-forcing precoding. The precoding vectors are given as ${\mathbf{g}}_{k} = \zeta \hat {\mathbf{h}}_{k}$, where $\zeta $ satisfies the power constraint  $ \mathbb{E}[||\mathbf{x}||^{2}] \leq  P_{max}$ as $\zeta^2 = P_{max}/\Psi$, where $\Psi =\mathbb{E}\left[{\rm{tr}\left( {\mathbf{P}\hat {\mathbf{H}}}{ \hat {\mathbf{H}}^H} \right)}\right]$ and $\hat{\mathbf{H}}^H=[\hat {\mathbf{h}}_{1}, \hat {\mathbf{h}}_{2} \dots \hat {\mathbf{h}}_{K}]\in \mathbb{C}^{M\times K}$.

Our analysis is based on an ergodic achievable net-rate expression  from \cite{HJY}, that was derived exploiting the channel hardening property of large-scale multiple~input~multiple~output (MIMO) systems for which asymptotically is sufficient for each user to only know $\mathbb{E}[\mathbf{h}_{k}^{H} \mathbf{g}_{k}]$.  Assuming that $\mathbb{E}[\mathbf{h}_{k}^{H} \mathbf{g}_{k}]$ is perfectly known at user $k$, and treating  interference and channel gain uncertainty as worst-case independent Gaussian noise, it can be shown that user $k$ can achieve the ergodic net rate~\cite{HJY} \vspace{-.1in}
\begin{align}
\label{rate_11}
R_{k}=\left(1-\frac{S \tau_S}{\tau_C}  \right) \log_2(1+\gamma_{k}),
\end{align}
where $\tau_C$ is the length of each coherence block, and $\gamma_k$ is the downlink SINR of user $k$ given  under MRT precoding as \vspace{-.05in}
\begin{align}
\label{SINR_MRT}
& \gamma_{k}=\frac{p_{k} |\mathbb{E}[\mathbf{h}_{k}^{H} \hat{\mathbf{h}}_{k}]|^{2}}{p_{k} \mathbb{V}\text{ar}[\mathbf{h}_{k}^{H} \hat{\mathbf{h}}_{k}] + \sum_{f\neq k} p_{f} \mathbb{E} [|\mathbf{h}_{k}^{H} \hat{\mathbf{h}}_{f}|^{2}]+ \frac{\Psi}{\rho}},
\end{align}
where $\rho=\frac{P_{max}}{\sigma^2}$. The ergodic   net sum-rate is  given as \vspace{-.1in}
\begin{equation}
\label{R_sum_MC}
R_{sum}= \sum_{k=1}^{K}\left(1-\frac{S \tau_S}{\tau_C}  \right) \log_2(1+{\gamma}_{k}).
\end{equation}

Our goal is to analyze  \eqref{R_sum_MC}  in the large system limit under the two considered CE protocols. The analysis will be done with the objectives of (i) developing optimized RISs designs under different CSI assumptions, and (ii) comparing the net sum-rate performance under   different CSI assumptions.

\section{Main Results}
We now present the asymptotic expressions of the net sum-rate and use them to optimize the RISs phase-shifts.

\subsection{Asymptotic Analysis}

 While the users' ergodic rates in \eqref{rate_11} are generally difficult to study for finite system dimensions, they  tend to approach deterministic quantities as the system dimensions grow large. These deterministic equivalents are  very accurate for moderate system dimensions as well  \cite{ annie,HJY}, and only depend on the channel statistics which facilitates formulating and solving optimization problems based on S-CSI.  Under this motivation, we compute the deterministic approximations of the users' rates under the following assumptions  \cite{HJY, annie}.

\textit{ Assumption 1.}
$M$, $N$ and $K$ grow large with a bounded ratio as $0< \liminf_{M,K \rightarrow \infty} \frac{K}{M}\leq \limsup_{M,K \rightarrow \infty}\frac{K}{M}<\infty$ and $0< \liminf_{M,N \rightarrow \infty} \frac{M}{N}\leq \limsup_{M,N \rightarrow \infty}\frac{M}{N}<\infty$. 

\textit{ Assumption 2.}
 $\mathbf{H}_1$ satisfies $\limsup\limits_{M,N\rightarrow \infty} ||\mathbf{H}_1\mathbf{H}_1^{H}||<\infty$.




\begin{theorem} \label{thm1_ric} Under Assumptions 1 and 2, the SINR of user $k$  in (\ref{SINR_MRT}) for the channel in (\ref{eq_ch1_RIC}) and its estimate in (\ref{est_corr_RIC}) under the MMSE-DFT CE  protocol satisfies $\gamma_{k}-{\gamma}_{k}^{\circ }\xrightarrow[M,N,K\rightarrow \infty]{a.s} 0$,  where $\gamma_k^{\circ}$ is given in \eqref{det_SINR_RIC},  $\mathbf{D}_{k}=\bar{\mathbf{h}}_{dk}\bar{\mathbf{h}}^{H}_{dk}+\bar{\mathbf{h}}_{dk}\sum_{l=1}^{L}\bar{\mathbf{h}}_{2lk}^{{H}} \boldsymbol{\Theta}_l^H \mathbf{H}_{1l}^H+\sum_{l=1}^{L}\mathbf{H}_{1l} \boldsymbol{\Theta}_l \bar{\mathbf{h}}_{2lk} \bar{\mathbf{h}}_{dk}^{H}+\sum_{l=1}^{L}\sum_{l'=1}^{L}\mathbf{H}_{1l} \boldsymbol{\Theta}_l \bar{\mathbf{h}}_{2lk}\bar{\mathbf{h}}_{2l'k}^{{H}} \boldsymbol{\Theta}_{l'}^H \mathbf{H}_{1l'}^H$, $\mathbf{A}_{k}$ is defined under \eqref{eq_ch1_RIC}, and $\textbf{C}_{k}$ is defined in Lemma \ref{L1_RIC}.
\end{theorem}
\begin{IEEEproof}
The proof  follows by applying the asymptotic results in \cite[Appendix A]{HJY}.
\end{IEEEproof}

\begin{figure*}[!t]
\normalsize
\begin{align}
\label{det_SINR_RIC}
&{\gamma}_{k}^{\circ}=\frac{p_k\left|\frac{1}{M}\text{tr}\left(\mathbf{D}_{k}+\sum_{l=1}^{L}\frac{\beta_{2lk}^{{n}^2}}{\beta_{2lk}^{\rm n}+\frac{1}{S\rho_{p}\tau_S M \beta_{1l}}} \mathbf{H}_{1l}\mathbf{H}_{1l}^H+ \frac{\beta_{dk}^{{n}^2}}{\beta_{dk}^{\rm n}+\frac{1}{S\rho_{p}\tau_S}} \boldsymbol{I}_{M}\right)\right|^{2}}{\frac{1}{M}\sum_{f\neq k}{\frac{p_{f}}{M}\text{tr}((\mathbf{D}_{f}+\textbf{C}_{f})(\mathbf{D}_{k}+\mathbf{A}_{k})})+\frac{\frac{p_{k}}{M}\sum_{k=1}^{K}{\frac{1}{M}\text{tr}(\mathbf{D}_{k}+\mathbf{C}_{k})}}{\rho}},
\end{align}
\begin{align}
\label{det_SINR_DE_ric}
&\gamma_{k}^{\circ}=\frac{p_{k}|\frac{1}{M}\text{tr}(\mathbf{D}_{k}+\mathbf{R}_{k} \mathbf{Q}_{k}\mathbf{R}_{k})|^2}{\frac{1}{M}\sum_{f\neq k} \frac{p_{f}}{M}\text{tr}((\mathbf{D}_{f}+\mathbf{R}_{f} \mathbf{Q}_{f}\mathbf{R}_{f})( \mathbf{D}_{k}+\mathbf{R}_{k}))+\frac{1}{M} \sum_{k=1}^{K}\frac{\frac{p_{k}}{M} \text{tr}(\mathbf{D}_{k}+\mathbf{R}_{k} \mathbf{Q}_{k} \mathbf{R}_{k})}{\rho}},
\end{align} 
\setcounter{equation}{13}
\begin{align}
\label{det_SINR_DE_ricwo}
&\gamma_{k}^{\circ}=\frac{p_{k}\left|\frac{1}{M}\left(\bar{\mathbf{h}}_{dk}^H \bar{\mathbf{h}}_{dk} +\frac{\beta_{dk}^{{n}^2}}{\beta_{dk}^{\rm n}+\frac{1}{\rho_{p}\tau_S}}M\right)\right|^2}{\frac{1}{M}\sum_{f\neq k} \frac{p_{f}}{M}\text{tr}\left(\left(\bar{\mathbf{h}}_{df} \bar{\mathbf{h}}^H_{df} +\frac{\beta_{df}^{{n}^2}\mathbf{I}_M}{\beta_{df}^{\rm n}+\frac{1}{\rho_{p}\tau_S}}\right)\left(\bar{\mathbf{h}}_{dk} \bar{\mathbf{h}}^H_{dk} +\frac{\beta_{dk}^{{n}^2}\mathbf{I}_M}{\beta_{dk}^{\rm n}+\frac{1}{\rho_{p}\tau_S}}\right)\right)+\frac{1}{M} \sum_{k=1}^{K}\frac{p_{k}}{M\rho}\left(\bar{\mathbf{h}}_{dk}^H \bar{\mathbf{h}}_{dk} +\frac{\beta_{dk}^{{n}^2}M}{\beta_{dk}^{\rm n}+\frac{1}{\rho_{p}\tau_S}}\right)}.
\end{align}
\hrulefill
\vspace*{4pt}
\end{figure*}


\begin{theorem}\label{thm2_ric} 
  Under Assumptions 1 and 2, the SINR of user $k$  in (\ref{SINR_MRT}) for the channel in (\ref{eq_ch1_RIC}) and its estimate in \eqref{est_deric} under the MMSE-DE CE protocol, satisfies  $\gamma_{k}-{\gamma}_{k}^{\circ}\xrightarrow[M,N,K\rightarrow \infty]{a.s} 0$, 	where  $\gamma^{\circ}_k$ is given by \eqref{det_SINR_DE_ric},  $\mathbf{R}_{k}$ and $\mathbf{Q}_{k}$ are defined in Lemma \ref{L4ric}, and $\mathbf{D}_k$ is defined in Theorem \ref{thm1_ric}.
\end{theorem}
\begin{IEEEproof}
The proof follows by using the asymptotic results in \cite{HJY}, while considering  estimates  in Lemma \ref{L4ric}.
\end{IEEEproof}


\begin{corollary}\label{cor_perric} Under the setting of Theorem \ref{thm1_ric} and assuming perfect CSI , $\gamma_k^{{\circ}}=\frac{p_{k}|\frac{1}{M}\text{tr}(\mathbf{D}_{k}+\mathbf{A}_{k})|^{2}}{\frac{1}{M}\sum_{f\neq k}\frac{p_{f}}{M}{\text{tr}((\mathbf{D}_{f}+\mathbf{A}_{f})(\mathbf{D}_{k}+\mathbf{A}_{k}}))+\frac{p_{k}}{M\rho}\sum_{k=1}^{K}{\frac{1}{M}\text{tr}(\mathbf{D}_{k}+\mathbf{A}_{k})}}$.
 \end{corollary}
\begin{IEEEproof}
The proof follows by letting $\rho_{p}\rightarrow \infty$ in \eqref{det_SINR_RIC}.
\end{IEEEproof}

\begin{theorem}\label{thm2_ricwo} 
  Consider the setting of Theorem \ref{thm2_ric} without RISs, then ${\gamma}_{k}^{\circ}$ is given as \eqref{det_SINR_DE_ricwo}.
\end{theorem}
\begin{IEEEproof}
The proof  follows by setting $L=0$ in \eqref{det_SINR_DE_ric}.
\end{IEEEproof}

\begin{corollary} \label{Cor_rate_RIC}
Under Assumptions 1 and 2,  the users' ergodic achievable net rates in \eqref{rate_11} converge as $R_{k}-R_{k}^{\circ}\xrightarrow[M,N,K\rightarrow \infty]{a.s} 0$,  where $R_{k}^{\circ}= \left(1-\frac{S \tau_S}{\tau_C}  \right)\log(1+{\gamma}_{k}^{\circ})$ and ${\gamma}_{k}^{\circ}$ is given by (\ref{det_SINR_RIC}) with $S=\frac{NL}{M}+1$ for the MMSE-DFT CE protocol, or given by \eqref{det_SINR_DE_ric} with $S=1$ for the MMSE-DE CE protocol.
\end{corollary}
\begin{IEEEproof}
The proof follows by applying the continuous mapping theorem  on $R_k^{\rm ric}$.
\end{IEEEproof}

 An asymptotic approximation for the ergodic achievable  net sum-rate can be obtained as 
 \setcounter{equation}{14}
\begin{equation}
\label{R_sum_ric}
R_{sum}^{\circ}= \sum_{k=1}^{K}\left(1-\frac{S \tau_S}{\tau_C}  \right) \log(1+{\gamma}_{k}^{\circ}).
\end{equation}

 
The deterministic equivalents in \eqref{det_SINR_RIC}, \eqref{det_SINR_DE_ric} and \eqref{det_SINR_DE_ricwo} provide some useful insights. The desired signal energy in the numerator of  these expressions  stays constant with respect to $M$, while  the interference and noise terms in the denominators of all three expressions vanish  as $M\rightarrow \infty$ while $K, N$ and $L$ are kept fixed. Therefore, the SINR grows with $M$ for fixed $K$ \cite{HJY}. We also note from  \eqref{det_SINR_RIC} and \eqref{det_SINR_DE_ric} that asymptotically, the RIS matrices $\boldsymbol{\Theta}_l$'s only appear in  all terms involving the LoS channel components, and will therefore yield higher performance gains for large  Rician factors $\kappa$. Moreover, the desired signal  and interference  (first term in denominator) terms scale quadratically with $N$ and $L$, while the noise term scales linearly with $N$ and $L$, indicating more RIS gains in noise limited scenarios. However, we can optimize   phases to improve  SINR  in interference limited scenarios as we do next.

\subsection{Ergodic Net  Sum-Rate Maximization using S-CSI}

Next we optimize the RIS phase-shifts using knowledge of only channel statistics that characterize the derived deterministic equivalents. This S-CSI changes much slower than the actual fast fading channel, and therefore the  phase shifts need to be optimized only once after several coherence periods. The net sum-rate maximization problem is formulated below. \vspace{-.05in}
 \begin{subequations}
 \begin{alignat}{2} \textit{(P1)} \hspace{.15in}
&\!\max_{\boldsymbol{\phi}} \hspace{.25in} &  \sum_{k=1}^{K} \left(1-\frac{S \tau_S}{\tau_C}  \right)  \log(1+\gamma_{k}^{\circ}) \label{P1}\\
&\text{s.t.} & \hspace{.25in} |\phi_{ln}|=1, \hspace{.08in} \forall l,n\label{constraint7}
\end{alignat}
\end{subequations} 
where $\phi_{ln}$ is the $n^{th}$ diagonal element of $\boldsymbol{\Theta}_l$, and  $\boldsymbol{\phi}=[\phi_{11}, \dots, \phi_{1N}, \phi_{21}, \dots, \phi_{LN}]^T\in \mathbb{C}^{NL\times 1}$.


\begin{algorithm}[!t]
\caption{Projected Gradient Ascent Algorithm}\label{alg:euclid}
\begin{algorithmic}[1]
\State \textbf{Initialize:} $\boldsymbol{\phi}^1$, $R_{sum}^{{\circ^1}}=f(\boldsymbol{\phi}^1)$ where $f(.)$ is given by \eqref{R_sum_ric}, $\epsilon>0$, $s=1$.
\State  \textbf{Repeat}
\State $\bar{R}_{sum}^{{\circ}}=R_{sum}^{{\circ^s}}$;
\State $[\mathbf{p}^{s}]_{N(l-1)+n}=\frac{\partial R_{sum}^{\circ}}{\partial \phi_{ln}}|_{\boldsymbol{\phi}^s}$, $n=1,\dots, N$, $l=1,\dots, L$;
\State $\tilde{\boldsymbol{\phi}}^{s+1}=\boldsymbol{\phi}^{s}+\mu \mathbf{p}^{s}$; 
\State  $\boldsymbol{\phi}^{s+1}=\exp(j \text{arg } (\tilde{\boldsymbol{\phi}}^{s+1}))$;
\State $R_{sum}^{{\circ^{s+1}}}=f(\boldsymbol{\phi}^{s+1})$; Update $s=s+1$;
\State \textbf{Until} $||R_{sum}^{{\circ^{s}}}-\bar{R}_{sum}^{{\circ}}||^2< \epsilon$; \textbf{Output} $\boldsymbol{\phi}^*=\boldsymbol{\phi}^{s}$;
\end{algorithmic}
\end{algorithm}

\textit{(P1)} is a constrained maximization problem that can be solved using projected gradient ascent as outlined in Algorithm \ref{alg:euclid}. We increase the objective by  iteratively updating the  phase-shifts vector $\boldsymbol{\phi}^{s}$ at iteration $s$ in a step proportional to the positive gradient $\mathbf{p}^{s}$ as $\tilde{\boldsymbol{\phi}}^{s+1} =\boldsymbol{\phi}^{s}+\mu \mathbf{p}^{s}$, where $\mu$ is obtained using backtracking line search. The solution $\tilde{\boldsymbol{\phi}}^{s+1}$ is projected to the closest feasible point satisfying  (\ref{constraint7}) as $\boldsymbol{\phi}^{s+1}=\exp(j \text{arg } (\tilde{\boldsymbol{\phi}}^{s+1}))$ \cite{annie}. Note that gradient ascent   only provides a local optimal solution to \textit{(P1)}, but we verify the large gains yielded by  the proposed design in simulations.

  The derivative of $R_{sum}^{{\circ}}$, which is the objective in \eqref{P1},  with  $\gamma_k^{{\circ}}$ given by \eqref{det_SINR_RIC} under the MMSE-DFT CE protocol,  with respect to $\phi_{ln}$ is derived  in the extended version of this work in \cite[Sec. IV-A]{bayan_full}.  The derivative of $R_{sum}^{{\circ}}$  with  $\gamma_k^{{\circ}}$ given  by \eqref{det_SINR_DE_ric}  under the MMSE-DE CE protocol, with respect to $\phi_{ln}$ is also derived in \cite[Sec. IV-A]{bayan_full}.

\subsection{ Instantaneous Sum-Rate Maximization Using Full I-CSI}
Since we obtain the full I-CSI of all channels under the MMSE-DFT protocol, we formulate the instantaneous net sum-rate expression and  propose to  maximize it  in this section, as a performance benchmark  for the S-CSI based RISs design.  Since the BS only has the channel estimates, we can write $y_k$ in \eqref{y_k} under MRT  as $y_k= \zeta \sqrt{p_k} \hat{\mathbf{h}}_k^{H} \hat{\mathbf{h}}_k s_k+ \zeta \sum_{f\neq k}^K \sqrt{p_f} \hat{\mathbf{h}}_k^{H} \hat{\mathbf{h}}_f s_f+ \zeta \sum_{f=1}^K \sqrt{p_f} \tilde{\mathbf{h}}_k^{H} \hat{\mathbf{h}}_f s_f+n_k$, where $ \tilde{\mathbf{h}}_k$ is the estimation error distributed as $\tilde{\mathbf{h}}_k\sim \mathcal{CN}(\mathbf{0}, \tilde{\mathbf{C}}_k)$ where $\tilde{\mathbf{C}}_k=\mathbf{A}_k-\mathbf{C}_k$, and $\mathbf{A}_k$ and $\mathbf{C}_k $ are defined in \eqref{eq_ch1_RIC} and Lemma \ref{L1_RIC} respectively. Note that  $\hat{\mathbf{h}}_k$ and $\tilde{\mathbf{h}}_k$ are functions of  $\boldsymbol{\Theta}_l$'s. Treating the last two terms  in $y_k$ as uncorrelated effective noise and  assuming I-CSI to be available at  users, we obtain the following instantaneous net rate expression.

 \setcounter{equation}{16}
\begin{theorem}
An achievable instantaneous net rate expression for user $k$ under the MMSE-DFT CE protocol is $R^{\text{inst}}_{k}=\left(1-\frac{S \tau_S}{\tau_C}  \right) \log_2(1+\gamma^{\text{inst}}_{k})$, where    $\gamma^{\text{inst}}_{k}$ is  given as
\begin{align}
& \gamma^{\text{inst}}_{k}=\frac{p_{k} |\hat{\mathbf{h}}_{k}^{H} \hat{\mathbf{h}}_{k}|^{2}}{\sum_{f\neq k} p_{f} |\hat{\mathbf{h}}_{k}^{H} \hat{\mathbf{h}}_{f}|^{2}+ \sum_{f=1}^K p_{f} \hat{\mathbf{h}}_f^{H} \tilde{\mathbf{C}}_k \hat{\mathbf{h}}_f + \frac{\Psi^{\text{inst}}}{\rho}}, \nonumber
\end{align}
where $\Psi^{\text{inst}}={\rm{tr}\left( {\mathbf{P}\hat {\mathbf{H}}}{ \hat {\mathbf{H}}^{H}} \right)}$  and $\rho=\frac{P_{max}}{\sigma^2}$.
\end{theorem}
The instantaneous net sum-rate is then given as  \vspace{-.05in}
\begin{align}
\label{rate_suminst}
R^{\text{inst}}_{sum}=\sum_{k=1}^K \left(1-\frac{S \tau_S}{\tau_C}  \right) \log_2(1+\gamma^{\text{inst}}_{k}),
\end{align}

Next, we formulate the   net sum-rate maximization problem similar to  \textit{(P1)} with the objective in \eqref{rate_suminst}. The RIS phase shifts are designed to solve this problem using the genetic algorithm  \cite{bayan_full}. Note that this full I-CSI based RISs design is expected to yield high sum-rate compared to the S-CSI based RIS design as the phase shifts are designed for each channel realization.  However, it also requires full CSI of all channels, i.e. $\hat{\mathbf{h}}_{dk}$ and $\hat{\mathbf{h}}_{2lk}$'s, which imposes a large training overhead of $S=NL/M+1$ sub-phases. This would compromise the net sum-rate, as we will see in Sec. IV, because the training loss factor  $\left(1-\frac{S \tau_S}{\tau_C}  \right)$ in \eqref{rate_suminst} will reduce as $N$ increases. 


\section{Simulation Results}
We consider the  BS   to be deployed at $(0,0,0)$m along the z-axis, and  the $L$ RISs  and $K$ users  to be placed along  arcs of radius $250$m and $400$m respectively that span angles from $-30^{\circ}$ to $30^{\circ}$ with respect to the $y$-axis. We define $p_k=1/K$, $P_{max}=10$W, and  $\sigma^2=-94$dBm. The path loss model is  given as $\beta_{k}=\frac{C_0}{d^{\bar{\alpha}}}$, with $C_0=30$\rm{dB},  $\bar{\alpha}_{1l}=2$,  $\bar{\alpha}_{2lk}=2.8$ and $\bar{\alpha}_{dk}=3.5$. The Rician factor is calculated as $\kappa_{ek}= 13-0.03d_{ek}$, where $e \in \{d, 2l\}$ and $d_{ek}$ is the  link distance.

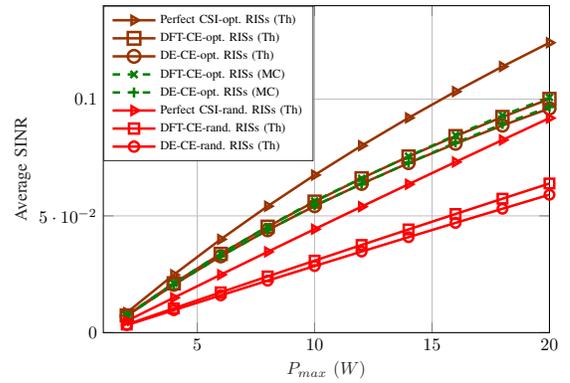
\begin{figure}[!t]
\centering
\tikzset{every picture/.style={scale=.95}, every node/.style={scale=.7}}
%
%
\definecolor{mycolor1}{rgb}{0.60000,0.20000,0.00000}%
\definecolor{mycolor2}{rgb}{0.00000,0.49804,0.00000}%
\begin{tikzpicture}

\begin{axis}[%
width=.75\columnwidth,
height=.55\columnwidth,
scale only axis,
xmin=1,
xmax=20,
xlabel style={font=\color{white!15!black}},
xlabel={$P_{max}$ $(W)$},
ymin=0,
ymax=0.14,
ylabel style={font=\color{white!15!black}},
ylabel={Average SINR},
axis background/.style={fill=white},
xmajorgrids,
ymajorgrids,
legend style={at={(axis cs: 1,0.14)},anchor=north west,legend cell align=left,align=left,draw=white!15!black, /tikz/column 2/.style={
                column sep=5pt,
            }},]
\addplot [color=mycolor1, line width=1.0pt,  mark=triangle, mark options={solid, rotate=270, mycolor1}]
  table[row sep=crcr]{%
2	0.00856584687550789\\
4	0.0247966910617771\\
6	0.0399482335224199\\
8	0.0541449002705827\\
10	0.0674910710785844\\
12	0.0800751817897544\\
14	0.0919728216011166\\
16	0.10324910843985\\
18	0.113960536410369\\
20	0.124156430881901\\
};
\addlegendentry{\scriptsize Perfect CSI-opt. RISs (Th)}

\addplot [color=mycolor1, line width=1.0pt,  mark=square, mark size=2.5pt, mark options={solid, mycolor1}]
  table[row sep=crcr]{%
2	0.00736995835550143\\
4	0.0211261391739996\\
6	0.0337317002481608\\
8	0.0453457070766677\\
10	0.0560970642256904\\
12	0.0660916867722567\\
14	0.0754176512788026\\
16	0.084148975312096\\
18	0.0923484431635327\\
20	0.100069753836228\\
};
\addlegendentry{\scriptsize  DFT-CE-opt. RISs (Th)}

\addplot [color=mycolor1, line width=1.0pt, mark=o,mark size=2.5pt, mark options={solid, mycolor1}]
  table[row sep=crcr]{%
2	0.0071776514775642\\
4	0.0205270809218494\\
6	0.0327066176231697\\
8	0.043884135899965\\
10	0.0541945969128743\\
12	0.0637481180606846\\
14	0.0726357034002355\\
16	0.0809334066497414\\
18	0.0887054163192941\\
20	0.0960063825449584\\
};
\addlegendentry{\scriptsize DE-CE-opt. RISs (Th)}

\addplot [color=mycolor2, dashed, line width=1.0pt,  mark=x, mark options={solid, mycolor2}]
  table[row sep=crcr]{%
2	0.00742223256720182\\
4	0.0211448283690985\\
6	0.0338130073467997\\
8	0.0454201604684503\\
10	0.0563967193607011\\
12	0.0660555132921863\\
14	0.0755187049507112\\
16	0.0845732892445153\\
18	0.0930317621556364\\
20	0.10072866213704\\
};
\addlegendentry{\scriptsize DFT-CE-opt. RISs (MC)}

\addplot [color=mycolor2, dashed, line width=1.0pt, mark=+, mark options={solid, mycolor2}]
  table[row sep=crcr]{%
2	0.00722128578025714\\
4	0.0205325538026032\\
6	0.0327428081972574\\
8	0.0439068692945081\\
10	0.0544161878206971\\
12	0.063628684321747\\
14	0.0726875766639627\\
16	0.0813461419611141\\
18	0.0893031917225323\\
20	0.0966492186452109\\
};
\addlegendentry{\scriptsize DE-CE-opt. RISs (MC)}

\addplot [color=red, line width=1.0pt,  mark=triangle, mark options={solid, rotate=270, red}]
  table[row sep=crcr]{%
2	0.00499781549553358\\
4	0.0149377593075945\\
6	0.0248041923234659\\
8	0.0345979877612694\\
10	0.0443200039404214\\
12	0.0539710846232396\\
14	0.0635520593465378\\
16	0.0730637437435694\\
18	0.0825069398566627\\
20	0.0918824364408801\\
};
\addlegendentry{\scriptsize Perfect CSI-rand. RISs (Th)}

\addplot [color=red, line width=1.0pt,  mark=square, mark options={solid, red}]
  table[row sep=crcr]{%
2	0.00347844758878455\\
4	0.0103949999512915\\
6	0.0172583530998494\\
8	0.0240692040833591\\
10	0.0308282361343473\\
12	0.0375361190486481\\
14	0.0441935095516164\\
16	0.0508010516514571\\
18	0.057359376980225\\
20	0.0638691051230199\\
};
\addlegendentry{\scriptsize DFT-CE-rand. RISs (Th)}

\addplot [color=red, line width=1.0pt,  mark=o, mark options={solid, red}]
  table[row sep=crcr]{%
2	0.00321876152411253\\
4	0.00961836798302375\\
6	0.0159679933521135\\
8	0.0222683142520297\\
10	0.0285199932652871\\
12	0.0347236793422093\\
14	0.0408800081917246\\
16	0.0469896026577022\\
18	0.0530530730814829\\
20	0.0590710176512216\\
};
\addlegendentry{\scriptsize DE-CE-rand. RISs (Th)}

\end{axis}

\end{tikzpicture}%
\caption{ SINR versus $P_{max}$ for  $M,N=60$ and $L,K=20$. } 
\label{Fig2a}
\end{figure}

\begin{figure}[!t]
\centering
\tikzset{every picture/.style={scale=.95}, every node/.style={scale=.7}}
%
%
\definecolor{mycolor1}{rgb}{0.60000,0.20000,0.00000}%
\definecolor{mycolor2}{rgb}{0.00000,0.49804,0.00000}%
\begin{tikzpicture}

\begin{axis}[%
width=.75\columnwidth,
height=.6\columnwidth,
scale only axis,
xmin=1,
xmax=20,
xlabel style={font=\color{white!15!black}},
xlabel={$P_{max}$ $(W)$},
ymin=0,
ymax=4.5,
ylabel style={font=\color{white!15!black}},
ylabel={Net sum rate (bps/Hz)},
axis background/.style={fill=white},
xmajorgrids,
ymajorgrids,
legend style={at={(axis cs:1,4.5)},anchor=north west,legend cell align=left,align=left,draw=white!15!black, /tikz/column 2/.style={
                column sep=5pt,
            }},]
            
            \addplot [color=mycolor1, line width=1.0pt, mark=triangle, mark options={solid, rotate=270, mycolor1}]
  table[row sep=crcr]{%
2	0.245478208632131\\
4	0.701899896033796\\
6	1.1185230499045\\
8	1.50135698780854\\
10	1.85514757649617\\
12	2.18371224736982\\
14	2.490169435789\\
16	2.77710045158634\\
18	3.04666658486738\\
20	3.300695671252\\
};
\addlegendentry{\scriptsize Perfect CSI-opt. RISs (Th)}

\addplot [color=mycolor1,  line width=1.0pt,  mark=square, mark size=2.5pt, mark options={solid, mycolor1}]
  table[row sep=crcr]{%
2	0.175721587390784\\
4	0.497627734797997\\
6	0.786253233815036\\
8	1.04729449094452\\
10	1.28511686789276\\
12	1.50314205054182\\
14	1.70410272579035\\
16	1.89021607839448\\
18	2.06330554766472\\
20	2.22488845421191\\
};
\addlegendentry{\scriptsize DFT-CE-opt. RISs (Th)}

\addplot [color=mycolor1,  line width=1.0pt,  mark=o, mark size=2.5pt,mark options={solid, mycolor1}]
  table[row sep=crcr]{%
2	0.204042387327688\\
4	0.576491540191524\\
6	0.909003960063153\\
8	1.20860290261588\\
10	1.48063220968529\\
12	1.72925513524674\\
14	1.95777914879811\\
16	2.16887539907998\\
18	2.36473164424908\\
20	2.54716163551654\\
};
\addlegendentry{\scriptsize DE-CE-opt. RISs (Th)}

\addplot [color=mycolor2, dashed, line width=1.0pt,  mark=x, mark options={solid, mycolor2}]
  table[row sep=crcr]{%
2	0.176952760147692\\
4	0.498059736378242\\
6	0.788019621992002\\
8	1.04898945950002\\
10	1.29199466294081\\
12	1.50228410867052\\
14	1.70632512852647\\
16	1.89940523762325\\
18	2.07744049017999\\
20	2.23847212952827\\
};
\addlegendentry{\scriptsize DFT-CE-opt. RISs (MC)}

\addplot [color=mycolor2, dashed, line width=1.0pt,  mark=+, mark options={solid, mycolor2}]
  table[row sep=crcr]{%
2	0.205267347929061\\
4	0.576644996483976\\
6	0.909959098899138\\
8	1.2092175538174\\
10	1.4867874897742\\
12	1.7262227729363\\
14	1.95937481317\\
16	2.17977865466275\\
18	2.37945338523826\\
20	2.56348693622406\\
};
\addlegendentry{\scriptsize DE-CE-opt. RISs (MC)}

\addplot [color=red, line width=1.0pt,  mark=triangle, mark options={solid, rotate=270, red}]
  table[row sep=crcr]{%
2	0.143838287894175\\
4	0.427746431543342\\
6	0.706753572510684\\
8	0.981002421525021\\
10	1.25062964427261\\
12	1.51576620290584\\
14	1.77653767322118\\
16	2.03306453960971\\
18	2.28546246966876\\
20	2.53384257017448\\
};
\addlegendentry{\scriptsize Perfect CSI-rand. RISs (Th)}

\addplot [color=red,  line width=1.0pt,  mark=square, mark options={solid, red}]
  table[row sep=crcr]{%
2	0.0833524256238262\\
4	0.248190048549565\\
6	0.410590430215234\\
8	0.570614396741367\\
10	0.728320538674703\\
12	0.883765322668987\\
14	1.03700319598229\\
16	1.18808668436209\\
18	1.33706648383512\\
20	1.48399154687033\\
};
\addlegendentry{\scriptsize DFT-CE-rand. RISs (Th)}

\addplot [color=red,  line width=1.0pt,  mark=o, mark options={solid, red}]
  table[row sep=crcr]{%
2	0.0919734719746468\\
4	0.273908939117882\\
6	0.453217760851156\\
8	0.62996419215052\\
10	0.804210160185253\\
12	0.976015379642115\\
14	1.14543746065381\\
16	1.31253200992026\\
18	1.47735272555524\\
20	1.63995148614071\\
};
\addlegendentry{\scriptsize DE-CE-rand. RISs (Th)}

\addplot [color=black, line width=1.0pt, mark=star, mark options={solid, black}]
  table[row sep=crcr]{%
2	0.101902158938797\\
4	0.303954237256323\\
6	0.503708216559567\\
8	0.701208816245816\\
10	0.896499491534371\\
12	1.08962248075902\\
14	1.28061885045159\\
16	1.46952853834049\\
18	1.65639039437996\\
20	1.84124221991832\\
};
\addlegendentry{\scriptsize No RISs-Perfect CSI}

\addplot [color=black, line width=1.0pt,  mark=diamond, mark options={solid, black}]
  table[row sep=crcr]{%
2	0.0565316385784911\\
4	0.168882780122322\\
6	0.280296635009907\\
8	0.390787570873041\\
10	0.500369616659857\\
12	0.609056473639134\\
14	0.716861525945939\\
16	0.823797850691992\\
18	0.929878227662664\\
20	1.03511514862125\\
};
\addlegendentry{\scriptsize No RISs-Imperfect CSI}

\end{axis}
\end{tikzpicture}%
\caption{Net sum-rate for \hspace{-.02in}$M,N=60$, and \hspace{-.01in}$L,K=20$.}
\label{Fig2b}
\end{figure}
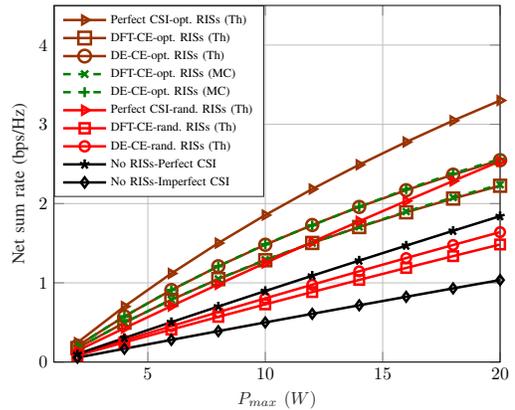

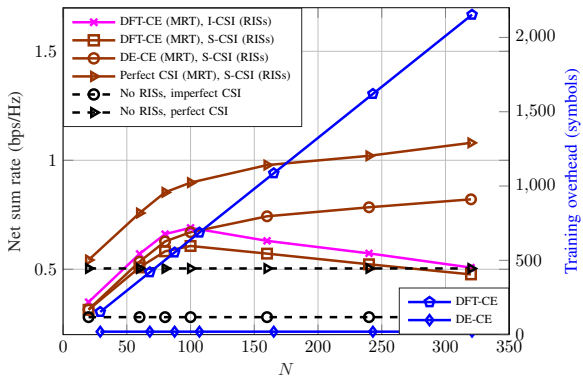
\begin{figure}[!t]
\centering
\tikzset{every picture/.style={scale=.95}, every node/.style={scale=.7}}
%
%
\definecolor{mycolor1}{rgb}{0.60000,0.20000,0.00000}%
\definecolor{mycolor2}{rgb}{1.00000,0.00000,1.00000}%
\begin{tikzpicture}

\begin{axis}[%
width=.75\columnwidth,
height=.55\columnwidth,
scale only axis,
xmin=0,
xmax=350,
xlabel style={font=\color{white!15!black}},
xlabel={$N$},
ymin=0.2,
ymax=1.7,
ylabel style={font=\color{white!15!black}},
ylabel={Net sum rate (bps/Hz)},
axis background/.style={fill=white},
xmajorgrids,
ymajorgrids,
legend style={at={(axis cs: 0,1.7)},anchor=north west,legend cell align=left,align=left,draw=white!15!black, /tikz/column 2/.style={
                column sep=5pt,
            }},]
\addplot [color=mycolor2, line width=1.0pt,mark=x, mark options={solid, mycolor2}]
  table[row sep=crcr]{%
20	0.349449235725966\\
60	0.570619686233413\\
80	0.661287749054878\\
100	0.690142889208297\\
160	0.630403836272409\\
240	0.572933374552736\\
320	0.507651962749354\\
};
\addlegendentry{\scriptsize DFT-CE (MRT), I-CSI (RISs)}

\addplot [color=mycolor1, line width=1.0pt,  mark=square, mark options={solid, mycolor1}]
  table[row sep=crcr]{%
20	0.314075851834456\\
60	0.508389231895292\\
80	0.582193661130019\\
100	0.606773646241047\\
160	0.571757693727538\\
240	0.52285484818227\\
320	0.476386131211849\\
};
\addlegendentry{\scriptsize DFT-CE (MRT), S-CSI (RISs)}

\addplot [color=mycolor1, line width=1.0pt, mark=o, mark options={solid, mycolor1}]
  table[row sep=crcr]{%
20	0.315403864875432\\
60	0.534703991773895\\
80	0.627650086401156\\
100	0.669878688293443\\
160	0.742971246432717\\
240	0.784401288512283\\
320	0.820857547318518\\
};
\addlegendentry{\scriptsize DE-CE (MRT), S-CSI (RISs)}

\addplot [color=mycolor1, line width=1.0pt, mark=triangle, mark options={solid, rotate=270, mycolor1}]
  table[row sep=crcr]{%
20	0.542652905048024\\
60	0.757955204618245\\
80	0.852926038367936\\
100	0.897578573839849\\
160	0.977993223514919\\
240	1.0205452093156334\\
320	1.08002031426666\\
};
\addlegendentry{\scriptsize Perfect CSI (MRT), S-CSI (RISs)}

\addplot [color=black, dashed, line width=1.0pt, mark=o, mark options={solid, black}]
  table[row sep=crcr]{%
20	0.280296635009907\\
60	0.280296635009907\\
80	0.280296635009907\\
100	0.280296635009907\\
160	0.280296635009907\\
240	0.280296635009907\\
320	0.280296635009907\\
};
\addlegendentry{\scriptsize No RISs, imperfect CSI}

\addplot [color=black, dashed, line width=1.0pt, mark=triangle, mark options={solid, rotate=270, black}]
  table[row sep=crcr]{%
20	0.503708216559567\\
60	0.503708216559567\\
80	0.503708216559567\\
100	0.503708216559567\\
160	0.503708216559567\\
240	0.503708216559567\\
320	0.503708216559567\\
};
\addlegendentry{\scriptsize No RISs, perfect CSI}

\end{axis}

\begin{axis}[
width=.75\columnwidth,
height=.55\columnwidth,
  axis y line*=right,
  axis x line=none,
	scale only axis,
  ymin=0, ymax=2200,
  ylabel=Training overhead (symbols),
  ylabel style={font=\color{blue}},
    legend style={at={(axis cs: 350,0)},anchor=south east,legend cell align=left,align=left,draw=white!15!black, /tikz/column 2/.style={
                column sep=5pt,
            }},]
]
\addplot [color=blue, line width=1.0pt, mark=pentagon, mark options={solid, blue}]
  table[row sep=crcr]{%
20	153\\
60	420\\
80	553\\
100	687\\
160	1086\\
240	1620\\
320	2153\\
};
\addlegendentry{\scriptsize  DFT-CE}
420        1220        1620        2020        3220        4820        6420
\addplot [color=blue, line width=1.0pt, mark=diamond, mark options={solid, blue}]
  table[row sep=crcr]{%
20	20\\
60	20\\
80	20\\
100	20\\
160	20\\
240	20\\
320	20\\
};
\addlegendentry{\scriptsize  DE-CE}

\end{axis}
\end{tikzpicture}%
\caption{Net sum-rate and training overhead  for $M=60$, $K=20$ and $L=20$. } 
\label{Fig5}
\end{figure}

We  first validate the tightness of the deterministic equivalents of the SINR   in Fig. \ref{Fig2a}.  The theoretical (Th) deterministic equivalents of the SINR under MMSE-DFT CE  in Theorem \ref{thm1_ric}, under MMSE-DE CE  in Theorem \ref{thm2_ric}, and under perfect CSI  in Corollary \ref{cor_perric} are plotted under random (rand.) as well as S-CSI based optimized (opt.) RISs phase shifts.      We also plot  the Monte-Carlo (MC) simulated SINR values in \eqref{SINR_MRT} under  both CE protocols considering S-CSI based RISs phase design. The figure shows an excellent match between the Monte-Carlo and theoretical SINR values even for moderate system sizes. We also observe that the SINR values are higher under MMSE-DFT CE protocol than those under DE protocol,  because the former achieves a  better estimation quality by using an optimal DFT based solution for the RIS phases during  CE. 



 
 In Fig. \ref{Fig2b}, we plot the ergodic net sum-rate in (\ref{R_sum_ric}) using the deterministic equivalent of the SINR in Theorem \ref{thm1_ric} for MMSE-DFT, the one in Theorem \ref{thm2_ric} for MMSE-DE, and the one in Corollary \ref{cor_perric} for perfect CSI, under random and optimized (Alg. 1) RISs phase shifts. We also plot the Monte-Carlo simulated net sum-rate in (\ref{R_sum_MC}) under  both CE protocols. In contrast to the observation from Fig. \ref{Fig2a}, we see here  that the net sum-rate under MMSE-DFT is lower than that under the MMSE-DE CE protocol. This is because  the higher SINR obtained under the MMSE-DFT CE protocol due to the better channel estimation quality comes at the expense of a large training overhead of $S\tau_S =(NL/M +1)K$ symbols to construct $\hat{\mathbf{h}}_k$'s using full CSI. On the other hand, the MMSE-DE protocol estimates all $\mathbf{h}_k$'s in just $\tau_S=K$ symbols.  We also observe that  the S-CSI based RISs  design performs significantly better compared to a system without RISs.



In Fig. \ref{Fig5} we compare the net sum-rate performance of an RISs-assisted system against $N$ for three scenarios: (i) full I-CSI based RISs design using the MMSE-DFT CE protocol in Sec. III-C, (ii) S-CSI based RISs design in Algorithm 1 while considering  MMSE-DFT protocol to implement MRT, and (iii) S-CSI based RISs design in Algorithm 1 while considering MMSE-DE protocol to implement MRT. Note that for scenario (i) the average instantaneous net sum-rate in \eqref{rate_suminst} is plotted, while for scenarios (ii) and (iii) the deterministic equivalents of the ergodic net sum-rate in \eqref{R_sum_ric} are plotted.  We observe that the net sum-rate under MMSE-DFT CE protocol increases until a certain point ($N=100$) and then starts to decrease. This is because after this point the increase in the training overhead given by $S\tau_S =(NL/M +1)K$ symbols, plotted in blue on the right y-axis, becomes dominant over the increase in SINR that comes with $N$. On the other hand the training overhead of the MMSE-DE CE protocol is always $\tau_S=K$ symbols as plotted in blue on the figure. This results in  scenario (iii) to perform better than scenario (ii) due to the significantly improved training loss factor $\left(1-\frac{S\tau_S}{\tau_C}\right)$ in \eqref{R_sum_ric}.

Next we compare the performance of S-CSI and I-CSI based RIS designs.  We observe that (i) shows a similar trend as (ii) with the net sum-rate first increasing and then decreasing with $N$ due to the large training overhead incurred by the MMSE-DFT CE protocol. However the net sum-rate is higher under (i) than that under (ii) because we are using full I-CSI to optimize the RISs phase shifts  to realize favorable instantaneous channels, instead of optimizing them only to realize favorable channel statistics as done in scenario (ii).  When comparing all three  scenarios, MMSE-DE+S-CSI based RISs design  outperforms the I-CSI based RISs design for $N>100$ due to the very low training overhead of DE scheme. 

\section{Conclusion}
In this work, we studied the  net sum-rate performance of a distributed RISs-assisted multi-user MISO   system under the DFT-CE and the DE-CE protocols. Considering imperfect I-CSI for  precoding at the BS, we derived the deterministic equivalents of the net sum-rate under  each CE protocol, and used them to optimize the RIS phase shifts based on S-CSI. As a benchmark, we also devised a scheme where the  phase shifts were instantaneously optimized using full I-CSI obtained using the DFT-CE protocol. Results showed that DE of the aggregate channel for precoding with S-CSI based RISs design  outperforms both DFT-CE based schemes, i.e. the one with RISs designed using S-CSI as well as using full I-CSI, due to the significantly lower training overhead of DE scheme. 

\bibliographystyle{IEEEtran}
\bibliography{bib}
\vspace{-.1in}
\end{document}